\documentclass{kapproc} % Computer Modern font calls

\usepackage{t1enc}

\usepackage{procps} 

\usepackage[dvips]{graphicx}

\upperandlowercase

\kluwerbib % will produce this kind of bibliography entry:

\usepackage{epsf}

\begin{document}

\articletitle{Star and cluster formation in extreme environments}

\author{Richard de Grijs} 
\affil{Department of Physics \& Astronomy, University of Sheffield,
Hicks Building, Hounsfield Road, Sheffield S3 7RH, UK}
\email{R.deGrijs@sheffield.ac.uk}

\begin{abstract}
Current empirical evidence on the star-formation processes in the
extreme, high-pressure environments induced by galaxy encounters
(mostly based on high-resolution {\sl Hubble Space Telescope}
observations) strongly suggests that star {\it cluster} formation is
an important and perhaps even the dominant mode of star formation in
such starburst events. The sizes, luminosities, and mass estimates of
the young massive star clusters (YMCs) are entirely consistent with
what is expected for young Milky Way-type globular clusters
(GCs). Recent evidence lends support to the scenario that GCs, which
were once thought to be the oldest building blocks of galaxies, are
still forming today. Here, I present a novel empirical approach to
assess the shape of the initial-to-current YMC mass functions, and
hence their possible survival chances for a Hubble time.
\end{abstract}


\section{Star clusters as starburst tracers}

The production of luminous, massive yet compact star clusters seems to
be a hallmark of the most intense star-forming episodes in galaxies.
Young massive star clusters (YMCs; with masses often significantly
exceeding $M_{\rm cl} = 10^5 {\rm M}_\odot$) are generally found in
intense starburst regions, often in galaxies involved in gravitational
interactions of some sort (e.g., de Grijs et al.  2001, 2003a,b,c,d,e
and references therein).

An increasingly large body of observational evidence suggests that a
large fraction of the star formation in starbursts actually takes
place in the form of such concentrated clusters, rather than in
small-scale star-forming ``pockets''. YMCs are therefore important as
benchmarks of cluster formation and evolution. They are also important
as tracers of the history of star formation of their host galaxies,
their chemical evolution, the initial mass function (IMF), and other
physical characteristics in starbursts.

Using optical observations of the ``Mice'' and ``Tadpole'' interacting
galaxies (NGC 4676 and UGC 10214, respectively) -- based on a subset
of the Early Release Observations obtained with the {\sl Advanced
Camera for Surveys} on board the {\sl Hubble Space Telescope (HST)} --
and the novel technique of pixel-by-pixel analysis of their
colour-colour and colour-magnitude diagrams, we deduced the systems'
star and star cluster formation histories (de Grijs et al. 2003e). In
both of these interacting systems we find several dozen YMCs (or,
alternatively, compact star-forming regions), which overlap spatially
with regions of active star formation in the galaxies' tidal tails and
spiral arms (from a comparison with H$\alpha$ observations that trace
active star formation; Hibbard \& van Gorkom 1996). The tidal tail of
the Tadpole system is dominated by star-forming regions, which
contribute $\sim 70$\% of the total flux in the {\sl HST} F814W filter
(decreasing to $\sim 40$\% in the F439W filter). If the encounter
occurs between unevenly matched, gas-rich galaxies then, as expected,
the effects of the gravitational interaction are much more pronounced
in the smaller galaxy. For instance, when we compare the impact of the
interaction as evidenced by star cluster formation between M82 (de
Grijs et al. 2001, 2003b,c) and M81 (Chandar et al.  2001), or the
star cluster formation history in M51 (Bik et al. 2003), which is
currently in the process of merging with the smaller spiral galaxy NGC
5194, the evidence for enhanced cluster formation in the larger galaxy
is minimal if at all detectable.

Nevertheless, we have shown that star cluster formation is a major
mode of {\it newly-induced} star formation in galactic interactions,
with $\ge 35$\% of the active star formation in encounters occurring
in YMCs (de Grijs et al. 2003e).

The question remains, however, whether or not at least a fraction of
the numerous compact YMCs seen in extragalactic starbursts, may be the
progenitors of GC-type objects. If we could settle this issue
convincingly, one way or the other, the implications of such a result
would have profound and far-reaching implications for a wide range of
astrophysical questions, including (but not limited to) our
understanding of the process of galaxy formation and assembly, and the
process and conditions required for star (cluster) formation. Because
of the lack of a statistically significant sample of similar nearby
objects, however, we need to resort to either statistical arguments or
to the painstaking approach of one-by-one studies of individual
objects in more distant galaxies.

\section{From YMC to old globular cluster?}

The present state-of-the-art teaches us that the sizes, luminosities,
and -- in several cases -- spectroscopic mass estimates of most
(young) extragalactic star cluster systems are fully consistent with
the expected properties of young Milky Way-type GC progenitors.

However, the postulated evolutionary connection between the recently
formed YMCs in intensely star-forming areas, and old GCs similar to
those in the Galaxy is still a contentious issue. The evolution and
survivability of YMCs depend crucially on the stellar IMF of their
constituent stars (cf. Smith \& Gallagher 2001): if the IMF is too
shallow, i.e., if the clusters are significantly depleted in low-mass
stars compared to (for instance) the solar neighbourhood, they will
disperse within a few (galactic) orbital periods, and likely within
about a billion years of their formation (e.g., Smith \& Gallagher
2001, Mengel et al. 2002). Ideally, one would need to obtain (i)
high-resolution spectroscopy of all clusters in a given cluster sample
in order to obtain dynamical mass estimates (we will assume, for the
purpose of the present discussion, that our YMCs are fully virialised)
and (ii) high-resolution imaging (e.g., with the {\sl HST}) to measure
their luminosities and sizes. However, individual YMC spectroscopy,
while feasible today with 8m-class telescopes for the nearest systems,
is very time-consuming, since observations of large numbers of
clusters are required to obtain statistically significant
results. Instead, one of the most important and most widely used
diagnostics, both to infer the star (cluster) formation history of a
given galaxy, and to constrain scenarios for its expected future
evolution, is the distribution of cluster luminosities, or --
alternatively -- their associated masses, commonly referred to as the
cluster luminosity and mass functions (CLF, CMF), respectively.

Starting with the seminal work by Elson \& Fall (1985) on the young
cluster system in the Large Magellanic Cloud (LMC; with ages $\le 2
\times 10^9$ yr), an ever increasing body of evidence, mostly obtained
with the {\sl HST}, seems to imply that the CLF of YMCs is well
described by a power law. On the other hand, for the old GC systems in
the local Universe, with ages $\ge 10$ Gyr, the CLF shape is well
established to be roughly lognormal (Whitmore et al. 1993, Harris
1996, 2001, Harris et al. 1998).

This type of observational evidence has led to the popular -- but thus
far mostly speculative -- theoretical prediction that not only a
power-law, but {\it any} initial CLF (and CMF) will be rapidly
transformed into a lognormal distribution (e.g., Elmegreen \& Efremov
1997, Gnedin \& Ostriker 1997, Ostriker \& Gnedin 1997, Fall \& Zhang
2001). We recently reported the first discovery of an approximately
lognormal CLF (and CMF) for the star clusters in M82's fossil
starburst region ``B'', formed roughly simultaneously in a pronounced
burst of cluster formation (de Grijs et al. 2003b; see also Goudfrooij
et al. 2004). This provides the very first sufficiently deep CLF (and
CMF) for a star cluster population at intermediate age (of $\sim 1$
Gyr), which thus serves as an important benchmark for theories of the
evolution of star cluster systems.

The CLF shape and characteristic luminosity of the M82 B cluster
system is nearly identical to that of the apparently universal CLFs of
the old GC systems in the local Universe. This is likely to remain
virtually unchanged for a Hubble time, if the currently most popular
cluster disruption models hold. With the very short characteristic
cluster disruption time-scale governing M82 B (de Grijs et al. 2003c),
its cluster mass distribution will evolve toward a higher
characteristic mass scale than that of the Galactic GC system by the
time it reaches a similar age. Thus, this evidence, combined with the
similar cluster sizes (de Grijs et al. 2001), lends strong support to
a scenario in which the current M82 B cluster population will
eventually evolve into a significantly depleted old Milky Way-type GC
system dominated by a small number of high-mass clusters (de Grijs et
al. 2003b). This implies that (metal-rich) GCs, which were once
thought to be the oldest building blocks of galaxies, are still
forming today in galaxy interactions and mergers.

\section{The $L_V-\sigma_0$ relation as a diagnostic tool}

We have recently started to explore a new, empirical approach to
assess the long-term survival chances of YMCs formed profusely in
intense starburst environments (de Grijs, Wilkinson \& Tadhunter
2005). The method hinges on the empirical relationship for old
Galactic and M31 GCs, which occupy tightly constrained loci in the
plane defined by their $V$-band luminosities, $L_V$ (or, equivalently,
absolute magnitudes, $M_V$) and central velocity dispersions,
$\sigma_0$ (Djorgovski et al. 1997, and references therein; see
Fig. 1).

\begin{figure}[h!]
\begin{center}
\epsfxsize=10cm
\epsfbox{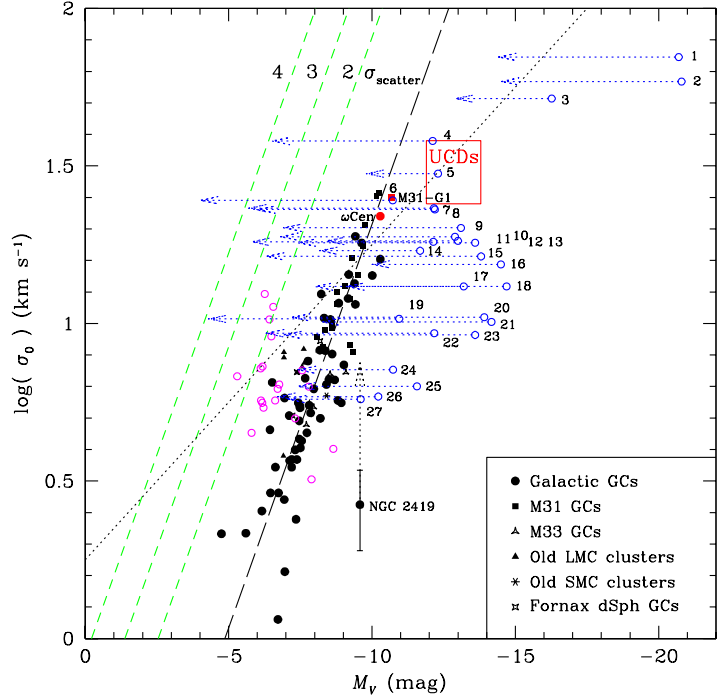}
\end{center}
\caption{Diagnostic figure used to predict the chances of survival to
old GC-type ages for YMCs with (central) velocity dispersion
measurements available in the literature. The filled symbols
correspond to the old GCs in the Local Group (see legend); the
best-fitting relation for these old clusters is shown by the
long-dashed line. The short-dashed lines are displaced from this
best-fitting relationship by, respectively, 2, 3, and 4 times the
scatter in the data points around the best-fitting line, $\sigma_{\rm
scatter}$. The dotted line is the Faber-Jackson relationship for
elliptical galaxies (see de Grijs et al. 2005), which bisects the
locus of the ultracompact dwarf galaxies (UCDs). The numbered open
circles are the locations of the YMCs with measured velocity
dispersions, which we have evolved to a common age of 12 Gyr
(represented by the dotted arrows) using the GALEV SSP models for the
appropriate metallicity and age of these objects. The remaining open
circles are the young compact clusters in the LMC and SMC. The most
massive GCs in both the Galaxy and M31 ($\omega$ Cen and G1,
respectively) are also indicated.}
\end{figure}

Encouraged by the tightness of the GC relationship, we also added the
available data points for the YMCs in the local Universe, including
nuclear star clusters (objects 1--5), for which velocity dispersion
information was readily available. In order to be able to compare them
to the ubiquitous old Local Group GCs, we evolved their luminosities
to a common age of 12 Gyr, adopting the ``standard'' Salpeter IMF
covering masses from 0.1 to 100 M$_\odot$, and assuming stellar
evolution as described by the GALEV SSPs (cf. Anders \& Fritze--v.
Alvensleben 2003). Based on a careful assessment of the uncertainties
associated with this luminosity evolution, we conclude that the most
important factor affecting the robustness of our conclusions is the
adopted form of the stellar IMF.

We find that if we adopt the universal solar neighbourhood IMF as the
basis for the YMCs' luminosity evolution, the large majority will
evolve to loci within twice the observational scatter around the
best-fitting GC relationship. In the absence of significant external
disturbances, this implies that these objects may potentially survive
to become old GC-type objects by the time they reach a similar
age. Thus, these results provide additional support to the suggestion
that the formation of proto-GCs appears to be continuing until the
present. Detailed one-to-one comparisons between our results based on
this new method with those obtained previously and independently based
on dynamical mass estimates and mass-to-light (M/L) ratio
considerations lend strong support to the feasibility and robustness
of our new method. The key characteristic and main advantage of this
method compared to the more complex analysis involved in using
dynamical mass estimates for this purpose is its simplicity and
empirical basis. Where dynamical mass estimates require one to obtain
accurate size estimates and to make assumptions regarding a system's
virialised state and M/L ratio, these complications can now be avoided
by using the empirically determined GC relationship as reference. The
only observables required are the system's (central or line-of-sight)
velocity dispersion and photometric properties.

Careful analysis of those YMCs that would overshoot the GC
relationship significantly if they were to survive for a Hubble time
show that their unusually high ambient density likely has already had
a devastating effect on their stellar content, despite their young
ages, thus altering their present-day mass function (PDMF) in a such a
way that they have become unstable to survive for any significant
length of time. This is, again, supported by independent analyses,
thus further strengthening the robustness of our new approach. The
expected loci in the $L_V-\sigma_0$ plane that these objects would
evolve to over a Hubble time are well beyond any GC luminosities for a
given velocity dispersion, thus leading us to conclude that they will
either dissolve long before reaching GC-type ages, or that they must
be characterised by a PDMF that is significantly depleted in low-mass
stars (or highly mass segregated). This, therefore, allows us to place
moderate limits on the functionality of their PDMFs.

\end{document}